# Comparing Disciplinary Classifications in SSH: Organizational, Channel-Based, and Text-Based Perspectives

Cristina Arhiliuc, Centre for R&D Monitoring (ECOOM), University of Antwerp, Antwerp, Belgium, cristina.arhiliuc@uantwerpen.be, ORCID: 0000-0002-5352-206X

Raf Guns, Anet Library Automation and Centre for R&D Monitoring (ECOOM), University of Antwerp, Antwerp, Belgium, raf.guns@uantwerpen.be, ORCID: 0000-0003-3129-0330

Tim C. E. Engels, Centre for R&D Monitoring (ECOOM), University of Antwerp, Antwerp, Belgium, tim.engels@uantwerpen.be, ORCID: 0000-0002-4869-7949

## Abstract

This study investigates how different approaches to disciplinary classification represent the Social Sciences and Humanities (SSH) in the Flemish VABB-SHW database. We compare organizational classification (based on author affiliation), channel-based cognitive classification (based on publication venues), and text-based publication-level classification (using channel titles, publication titles, and abstracts, depending on availability). The analysis shows that text-based classification generally aligns more closely with channel-based categories, confirming that the channel choice provides relevant information about publication content. At the same time, it is closer to organizational classification than channel-based categories are, suggesting that textual features capture author affiliations more directly than publishing channels do. Comparison across the three systems highlights cases of convergence and divergence, offering insights into how disciplines such as "Sociology" and "History" extend across fields, while "Law" remains more contained. Publication-level classification also clarifies the disciplinary profiles of multidisciplinary journals in the database, which in VABB-SHW show distinctive profiles with stronger emphases on SSH and health sciences. At the journal level, fewer than half of outlets with more than 50 publications have their channel-level classification fully or partially supported by more than 90% of publications. These results demonstrate the added value of text-based methods for validating classifications and for analysing disciplinary dynamics.

## Introduction

Efforts to classify scientific publications into disciplines support many aspects of research monitoring and evaluation. In the Social Sciences and Humanities (SSH), this task is especially complex due to diverse publication practices and overlapping disciplinary traditions (Archambault & Larivière, 2010; Sivertsen, 2016). Different approaches highlight different perspectives: organizational classifications reflect how research is structured within universities, channel-based classifications capture the disciplinary communities associated with publication venues, and text-based classifications derive information directly from the content of individual publications.

This study examines how these three perspectives relate to each other in the context of the Flemish Academic Bibliographic Database for the Social Sciences and Humanities (VABB-SHW). The database currently includes two classification systems. The organizational classification links each publication to the discipline(s) of the affiliated research unit(s) of the publication's authors and was designed to follow the research activity of SSH departments over time. Because the organizational classification mirrors how research is structured within universities, it is useful in local policy contexts. However, it does not take into account the content of individual publications and may be of limited relevance when scholars work across disciplinary boundaries. The channel-based cognitive classification (Guns et al., 2018) was introduced in 2017 and assigns publications channels (i.e. each journal, conference proceeding, and book) that occurs in the VABB-SHW[1] to disciplines. This approach helps to understand to which disciplinary communities Flemish SSH scholars contribute through the choice of their publication venue, and more broadly, it helps to understand the relative position of different SSH disciplines in Flanders. At the same time, it inherits well-known issues from external systems. These limitations also affect the VABB-SHW, where broad classifications at the level of the channel such as "Multidisciplinary" are not informative regarding the disciplinary coverage of individual contributions, and where reliance on external databases can lead to inconsistent labels.

To complement these systems and inform some of these limitations a text-based classification has recently been developed that uses a fully automated model to assign disciplines based on the title, abstract, and channel title of each publication (Arhiliuc et al., 2025). Unlike the organizational and channel-based classifications, the text-based classification is based directly on key parts of the content of a publication. This makes it a useful addition when studying publications in multidisciplinary channels or when exploring how different disciplines are represented in actual research output. At the same time, this method does not capture the broader context of how and why a publication was produced.

[1] The classification relies on a combination of international sources such as Web of Science, Scopus, OpenLibrary, ISSN.org, and DOAJ. The exact set of sources has evolved over time and is complemented with manual classification when needed.

This study is, to our knowledge, the first to conduct a systematic three-way comparison of organizational, channel-based, and text-based classifications for SSH publications. Previous research has primarily focused on pairwise comparisons. Guns et al. (2018) and Arhiliuc & Guns (2023), for example, examined the relationship between organizational and channel-based classification in SSH. Other studies have compared the classifications assigned at the level of channel (usually journals) and individual papers (often based on citation relationships), often with an emphasis on the sciences (Gong, 2023; Shu et al., 2019; Zhang & Shen, 2024). While these studies provide important insights into disciplinary assignment, no work has yet examined how all three perspectives – organizational, channel-based, and text-based – relate to each other, nor has such a comparison been conducted with a focus on SSH. To address this gap, we analyze the three classifications along three interrelated lines of inquiry:

- **Exploratory comparison of the three classifications**. How does the text-based classification relate to the channel-based and organizational classifications, and does it align more closely with one than with the other?
- **Analysis of publications where organizational and channel-based classifications diverge**. In these cases, does the text-based classification provide insight into whether authors are contributing knowledge from their organizational discipline(s) or from the disciplines of the channel? We expect the text-based classification in some cases to reflect aspects of both perspectives – for example, when a linguist publishes in a computer science journal, the content may draw from both fields.
- **Alignment between text-based classification and channel-level labels.** For publication channels with at least fifty entries, to what extent does the aggregated text-based classification of articles match the channel-level classification? In this part of the analysis, we look separately at multidisciplinary channels and at non-disciplinary channels. For the latter, we expect the text-based classification to broadly reflect the disciplinary label(s) assigned to the channel.

By bringing these three perspectives together, the study provides new insight into the strengths and limitations of different classification approaches in SSH. The comparison highlights where the classifications converge and diverge, and why. In doing so, it demonstrates the added value of a text-based approach as a complement to organizational and channel-based clasifications, and offers a more nuanced view of how disciplinary boundaries are reflected in SSH publication practices.

## Data

This study uses records from the 14th edition of the Flemish Academic Bibliographic Database for the Social Sciences and Humanities (VABB-SHW), which covers peer-reviewed publications authored by Flemish SSH scholars (Aspeslagh et al., 2024). The database was developed to support the national performance-based research funding system (PRFS) and to provide a reliable source for monitoring and studying SSH research output in Flanders (Verleysen et al., 2014).

The dataset includes 152,133 peer-reviewed publications from the period 2000–2022. For all but forty-eight of those records we have assigned discipline(s) under each of the three classifications considered in this study.

### Classification schemes

This study draws on three classification systems available in the VABB-SHW: the organizational classification, the channel-based cognitive classification, and a newly developed text-based classification.

Both the channel-based and text-based systems are structured around a modified version of the OECD Fields of Research and Development (FORD) classification. The FORD scheme organizes science into broad domains – referred to in this paper as "disciplinary areas" (e.g. Natural Sciences, Social Sciences, Humanities) – and into more specific subdomains, which we refer to as "disciplines". This scheme provides an international standard that facilitates comparison across contexts. In the VABB-SHW, the Humanities area is refined to increase granularity: disciplines that in the original FORD are grouped under broader categories are treated separately. In particular, instead of "History and Archaeology," the VABB-SHW distinguishes between "History" and "Archaeology"; instead of "Languages and Literature," it distinguishes between "Languages and Linguistics" and "Literature"; and instead of "Philosophy, Ethics and Religion," it distinguishes between "Philosophy and Ethics" and "Religion." In addition, the channel-based scheme contains one extra area – Multidisciplinary – to accommodate publication venues that cannot be assigned to a single field.

The organizational classification is structurally close to the modified version of OECD FORD used in the VABB-SHW, and most categories have a direct match. There are, however, a few notable exceptions. "Criminology" is treated as a separate discipline in the organizational scheme but is included under "Law" in FORD. "Social Health Sciences" belongs to “Social Sciences” in the organizational scheme, while their counterpart in FORD – “Health Sciences” – is part of the domain of Medical and Health Sciences. Finally, “Social and Economic Geography” has no direct counterpart in the organizational classification and is instead grouped under “Social Sciences (General).”

## Distribution of publications

The distribution of publications across disciplinary areas differs by classification type (Figure 1). By design, the organizational classification focuses on the SSH, as it was developed to monitor the output of researchers from SSH departments at Flemish universities. In contrast, the channel- and text-based classifications reflect researchers’ publication patterns and therefore extend beyond SSH, spreading into non-SSH disciplines as well.

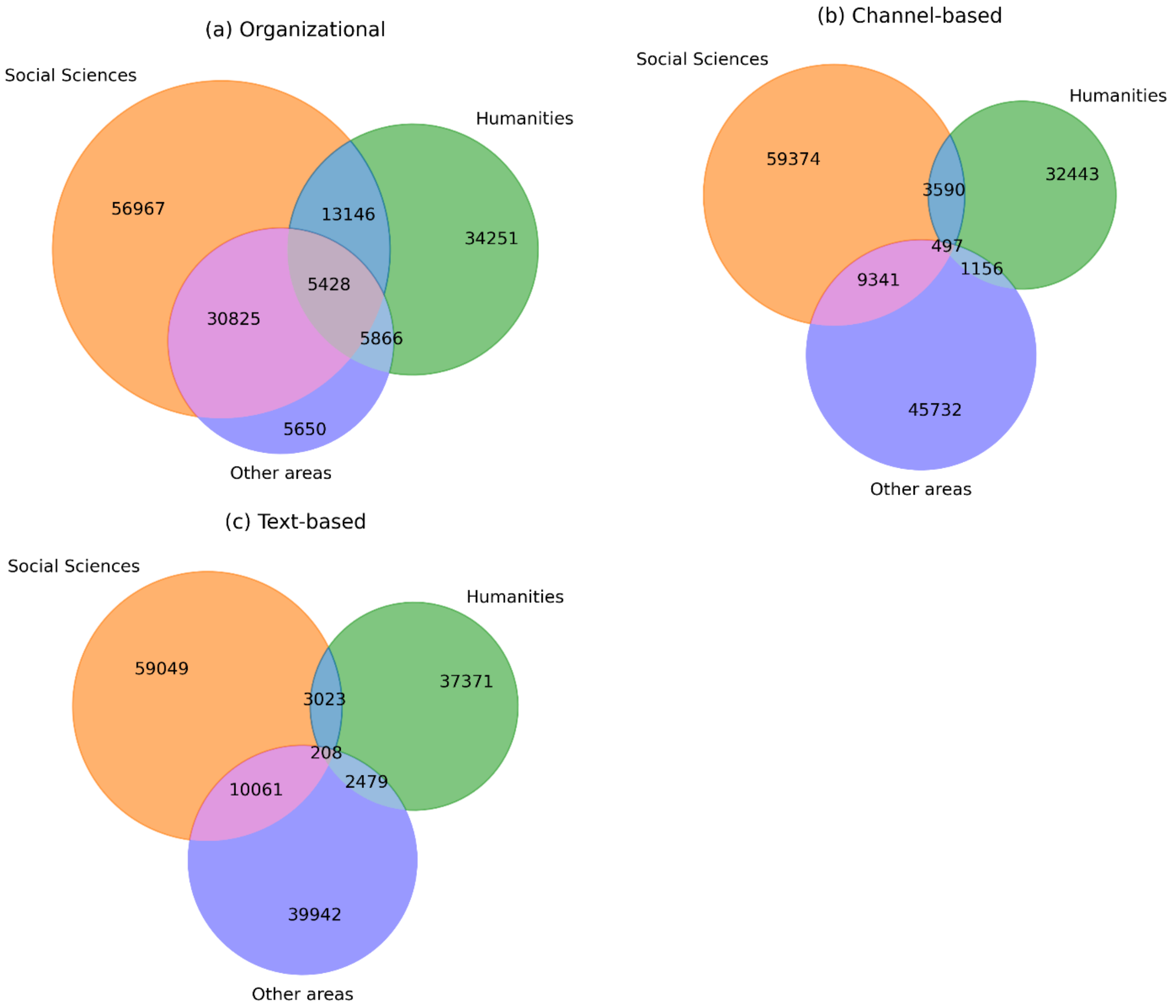

Figure 1. Venn diagram showing the distribution of publications across Social Sciences, Humanities, and other areas according to the three classification systems.

Beyond domain-level coverage, the three systems also differ in the number of disciplines attributed to individual publications (Figure 2). The organizational classification shows the broadest coverage, with many publications linked to two or more disciplines and a notable share exceeding three (6.54%). This is largely due to co-authorship across departmental boundaries, but also occurs in the cases where one researcher has multiple organizational affiliations. In contrast, the text-based classification is the most restrictive, with the majority of publications assigned to a single discipline. This reflects the characteristics of the training data, which was limited to a maximum of three disciplines and consisted primarily of single-discipline publications. The channel-based classification resembles the text-based in that most publications are assigned up to three disciplines. Although the manual component of the channel-based system is limited to three disciplines, the integration of external databases occasionally produces assignments beyond this threshold, but such cases are rare (1.15% of publications with more than three disciplines).

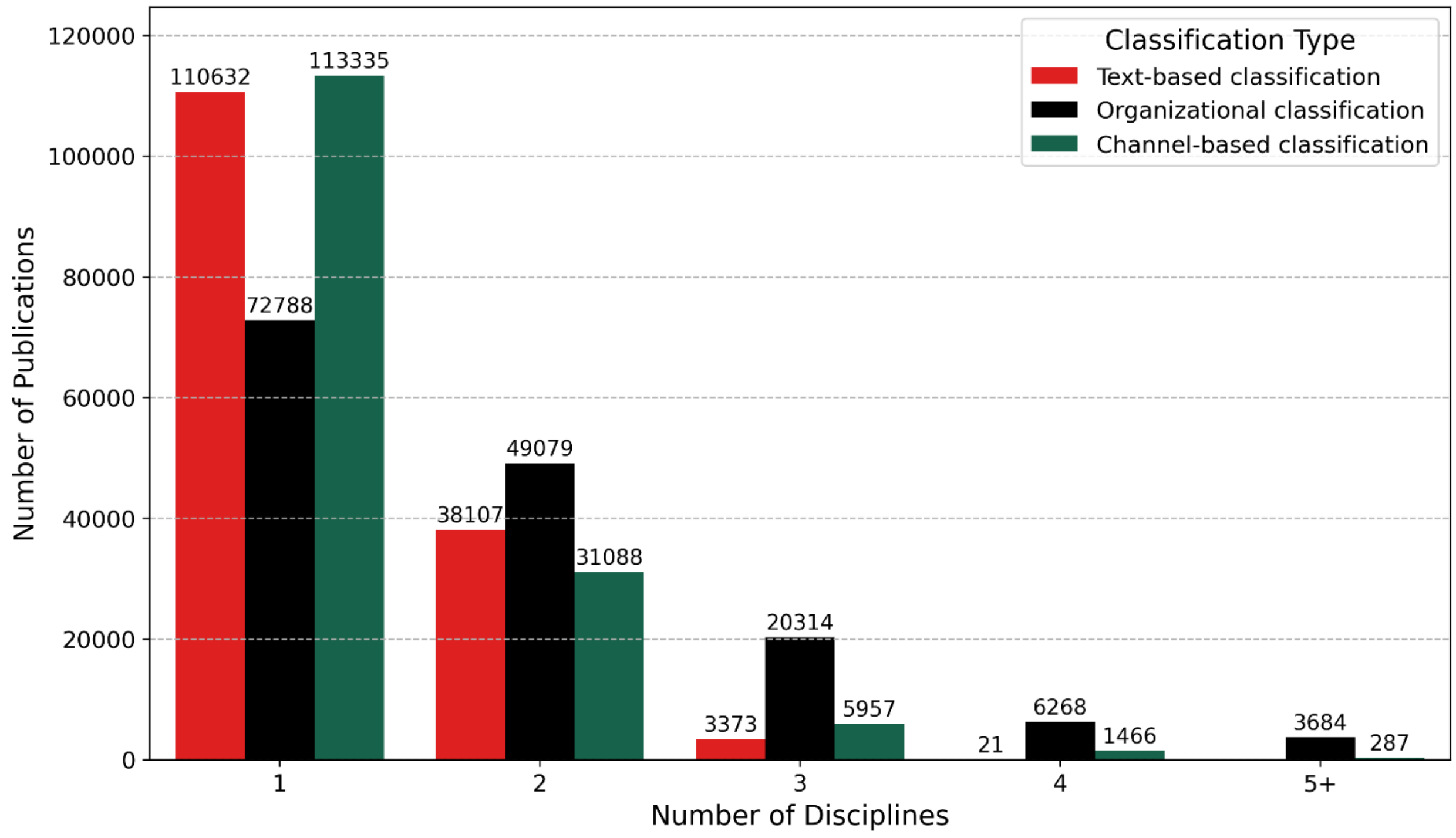

Figure 2. Distribution of publications by number of assigned disciplines under the three classification systems.

Together, these descriptive patterns highlight the similarities and differences between the three systems: while organizational classification captures SSH affiliation structures, the

channel- and text-based approaches follow publication venues and content, resulting in a broader spread across disciplines.

## Exploratory comparison of the three classifications

In this section, we explore how the text-based cognitive classification relates to the other two approaches. While earlier work (Guns et al., 2018) has compared the organizational and channel-based classifications, our analysis revisits this comparison to account for how the picture has evolved over time. The main emphasis, however, is on positioning the text-based results within this broader landscape of comparisons. In the following sections a publication assigned to multiple areas or disciplines contributes one count to each of its assigned categories.

### Organizational and channel-based classification

The relationship between organizational and channel-based classifications was previously analyzed using data up to 2015 (Guns et al., 2018). At that time, publications from Humanities departments were mainly classified in Humanities channels, with a smaller share extending into the other areas (0.73). Later work (Arhiliuc & Guns, 2023) confirmed that this picture had shifted, as the following years have seen a steady increase in the share of Humanities publications appearing in Social sciences channels. This development is also visible in the updated heatmap (Figure 3), where contributions from Humanities to Social sciences are much more pronounced than in earlier analyses. By contrast, the proportion of Social sciences publications classified in Social sciences channels has remained relatively stable over the same period.), where contributions from Humanities to Social sciences are much more pronounced than in earlier analyses. By contrast, the proportion of Social sciences publications classified in Social sciences channels has remained relatively stable over the same period.

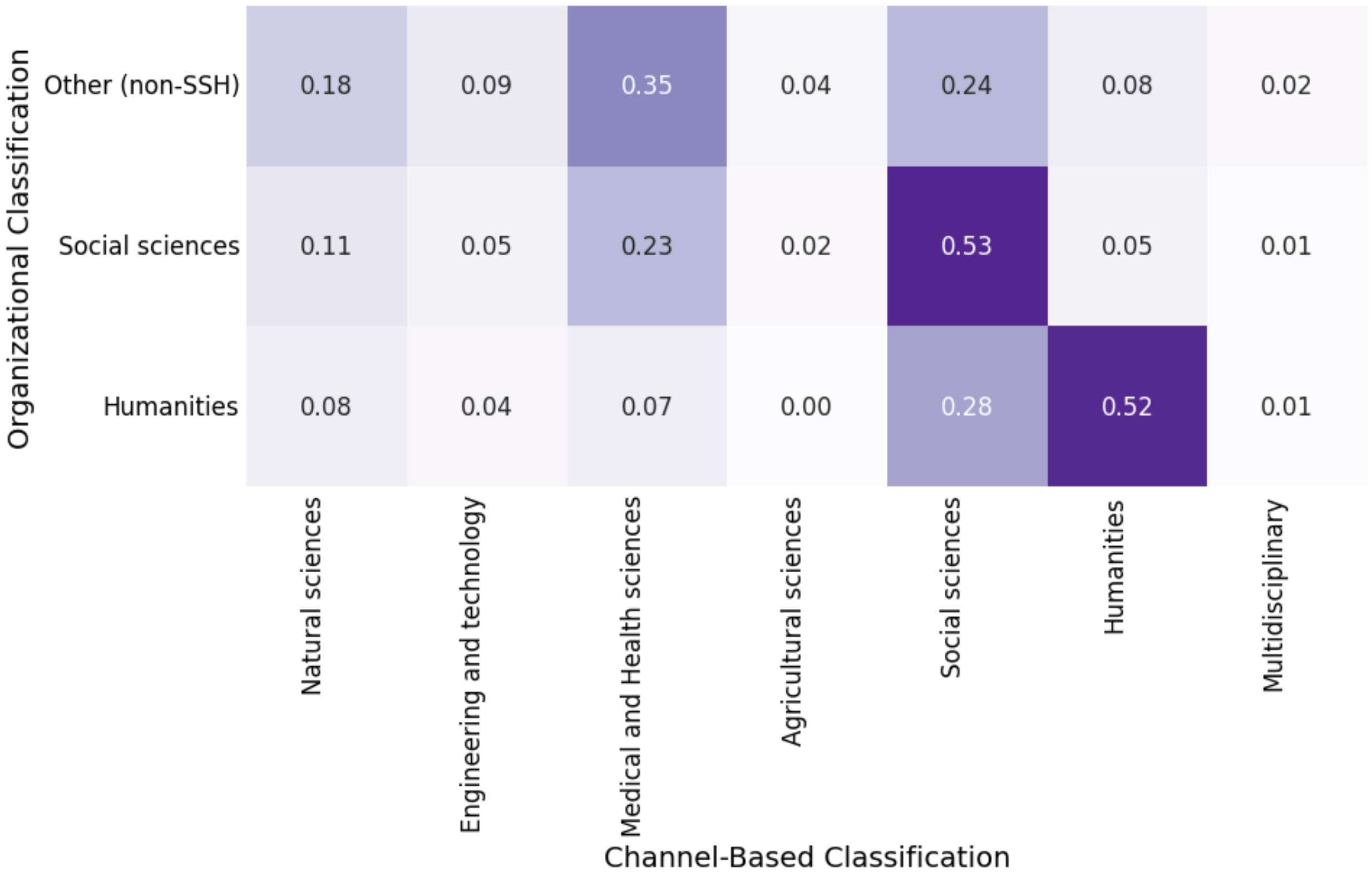

Figure 3 . Heatmap of overlap between organizational and channel-based classifications at the disciplinary area level, normalized per row.

## Organizational and text-based classifications

The comparison between organizational and text-based classifications provides insight into the extent to which the disciplinary orientation of researchers, defined by their departmental affiliations, corresponds to the content of their publications. Figure 4 presents the overlaps at the disciplinary area level, and Figure 5 at the discipline level, highlighting both areas of strong alignment and cases where researchers contribute beyond the boundaries of their home disciplines.

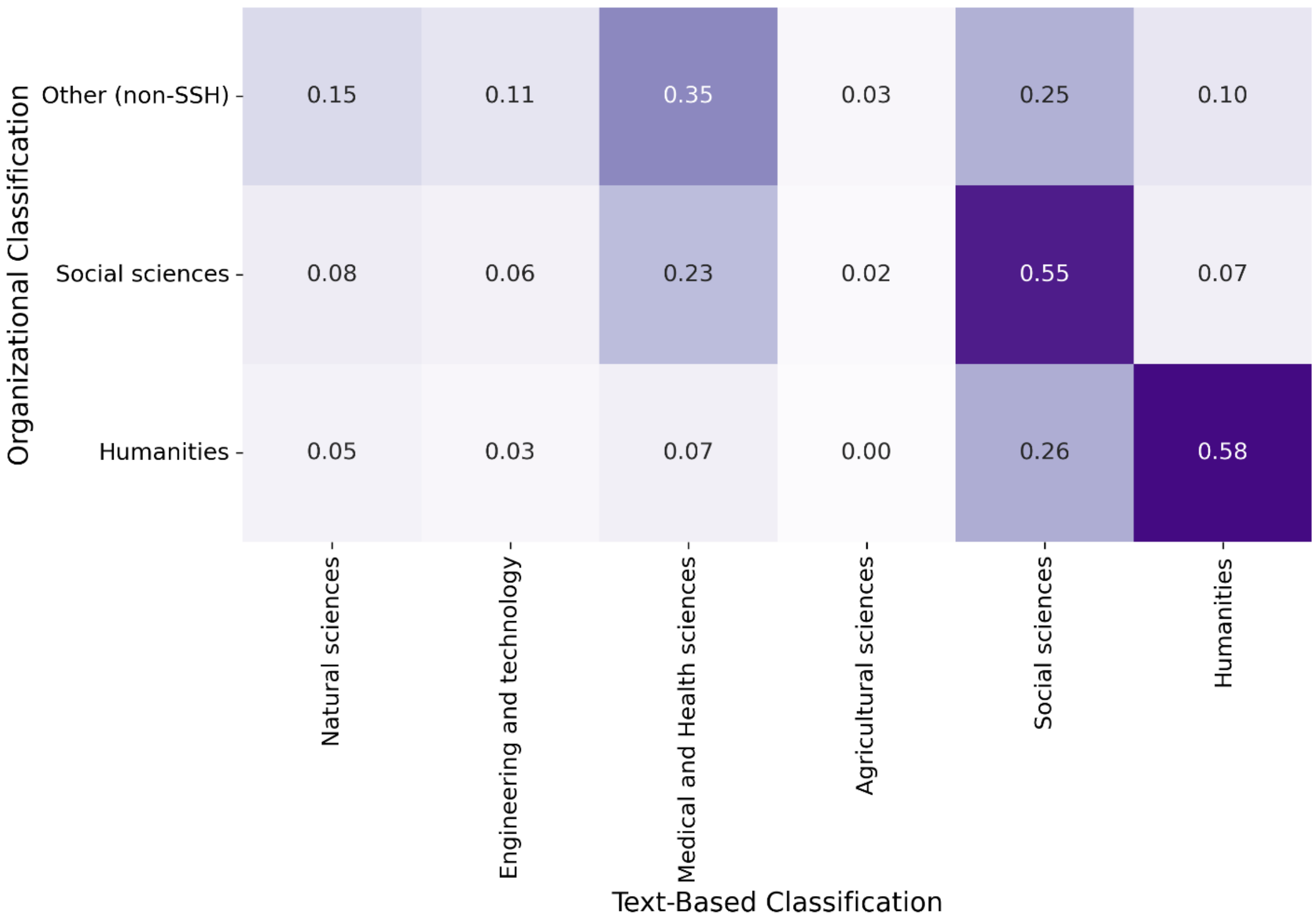

Figure 4. Heatmap of overlap between organizational and text-based classifications at the disciplinary area level, normalized per row.

At the disciplinary area level (Figure 4), publications classified through organizational affiliation align most strongly with their own disciplinary area in the text-based classification. For both Humanities and Social sciences, the largest share of publications is assigned to the same area, although the shares remain well below complete alignment (58% for Humanities and 55% for Social sciences). At the same time, there are notable contributions across areas: Humanities publications are frequently classified in the Social sciences, while Social sciences publications also extend into Medical and Health sciences.

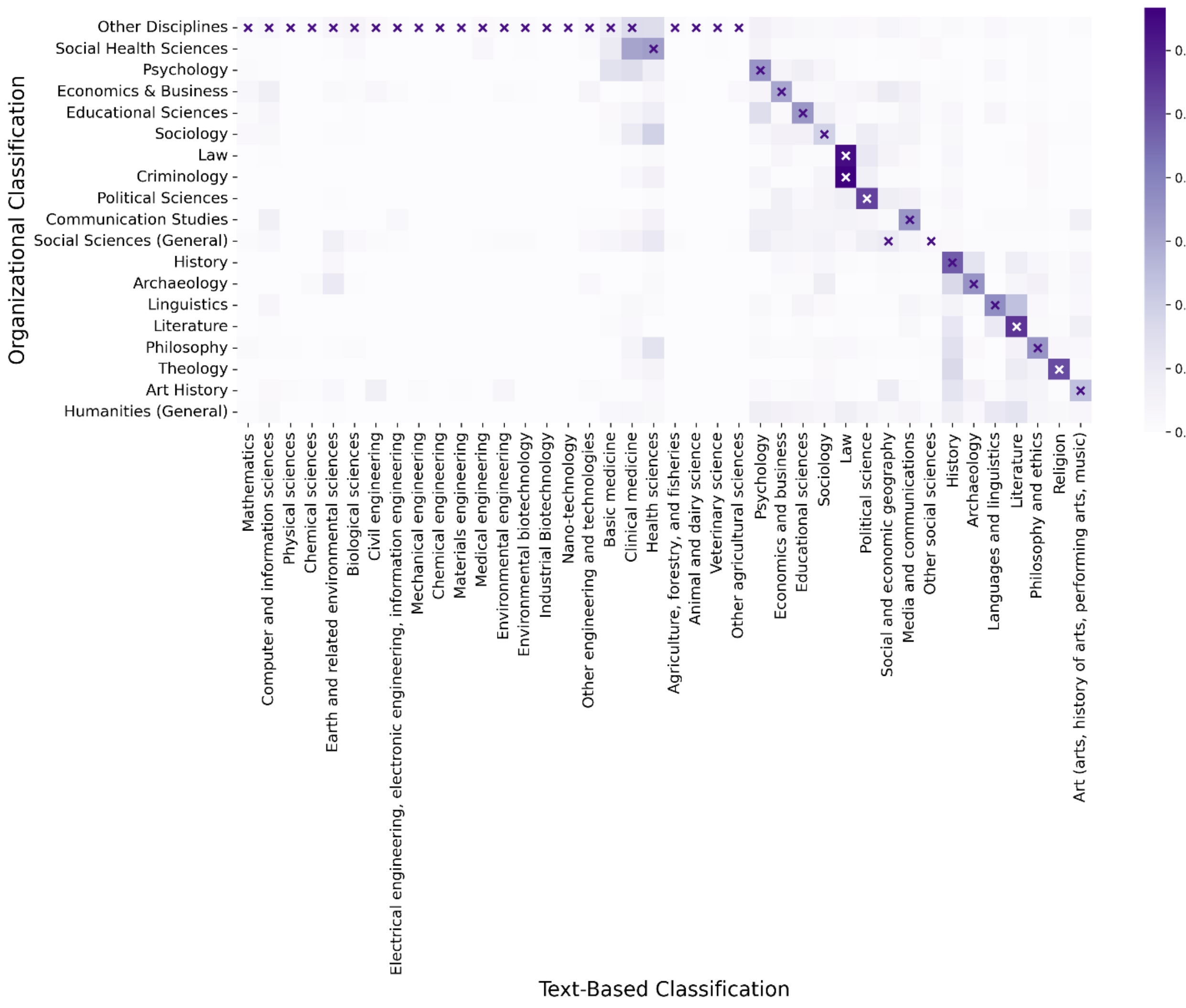

Figure 5. Heatmap of overlap between organizational and text-based classifications at the discipline level, normalized per row. The 'x' marks the disciplines mapped to each other.

At the discipline level (Figure 5), certain fields display strong agreement between the two classifications. "Law" is the clearest example, where publications from "Law" departments are consistently classified in the "Law" discipline by the text-based model. Other disciplines reveal more asymmetric relationships. For instance, publications from "Language and Linguistics" are frequently classified as "Literature", while "Literature" publications are less often classified as "Linguistics". This uneven relationship highlights how contributions flow more strongly in one direction than the other.

Finally, the text-based classification also underscores the cross-area reach of SSH disciplines. Social sciences show a notable extension into Medical and Health sciences, with contributions not only from "Social Health Sciences" but also from "Psychology" and "Sociology". Together, these patterns indicate that while organizational and text-based classifications align well in several areas, publications in SSH often extend beyond the boundaries of their home disciplines and disciplinary areas.

## Channel-based and text-based classifications

The comparison between channel-based and text-based classifications shows how publication channels relate to the content of outputs. Channel-based classification reflects the disciplinary orientation of venues, while text-based classification captures the profile of individual publications. Overlaps can therefore signal disciplinary contributions, either because a publication draws on another field or because channels in one area often include work from related fields. These results concern publications authored by SSH researchers in Flemish universities, which also explains the visibility of Social sciences, and to a lesser extent Humanities, across most areas.

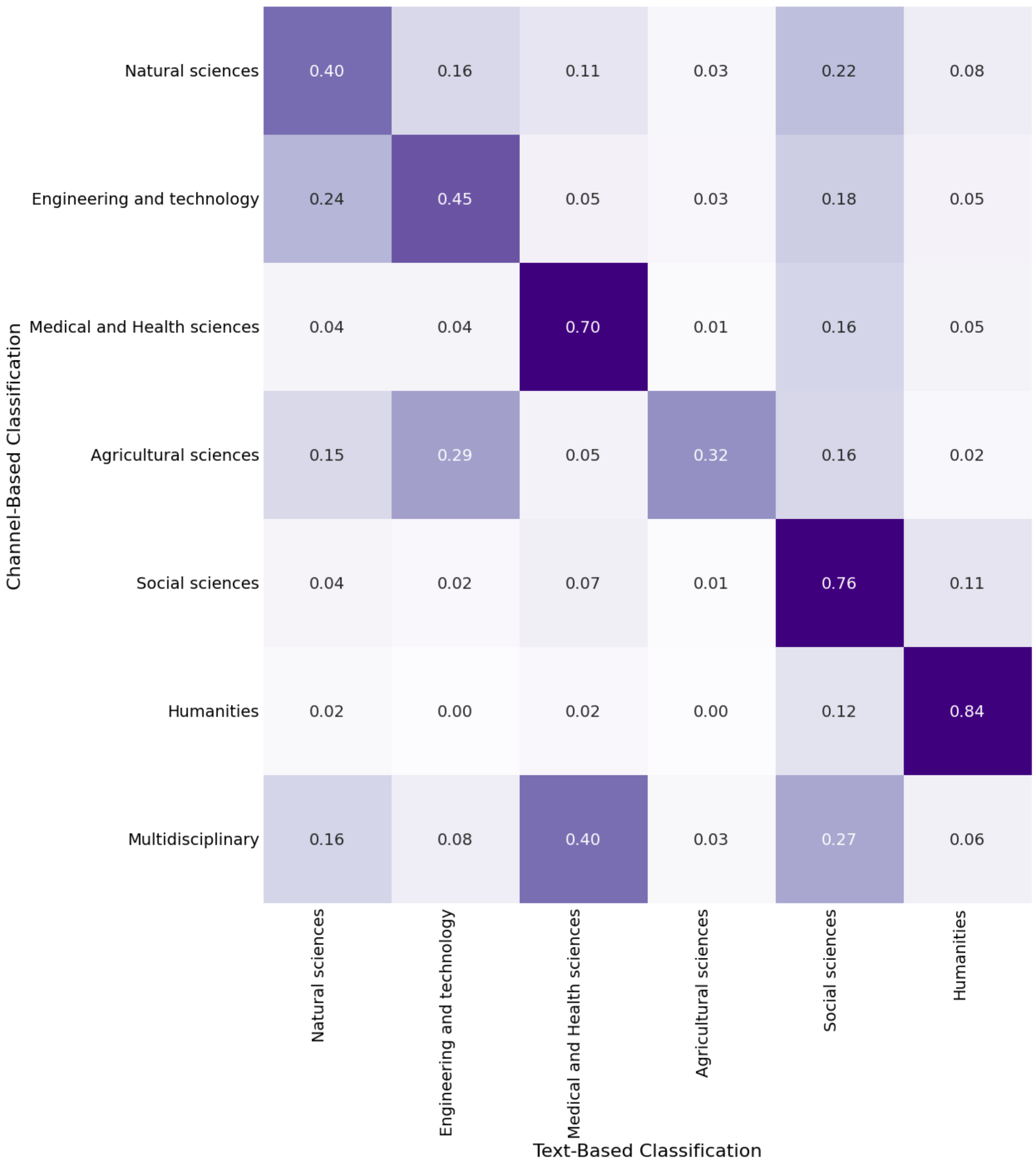

Figure 6. Heatmap of overlap between channel-based and text-based classifications at the disciplinary area level, normalized by row.

At the disciplinary area level (Figure 6), there is strong alignment between channel-based and text-based classifications in the Humanities, Social sciences, and Medical and Health sciences. By contrast, Natural sciences, Agricultural sciences, and Engineering show more dispersed patterns, with many publications classified into other areas. The category

Multidisciplinary also shows dispersion, but with a clear concentration in Medical and Health sciences and only very limited presence in Agricultural sciences, Humanities, and Engineering and technology. A consistent feature across nearly all rows is the presence of Social sciences, which reflects the composition of the dataset.

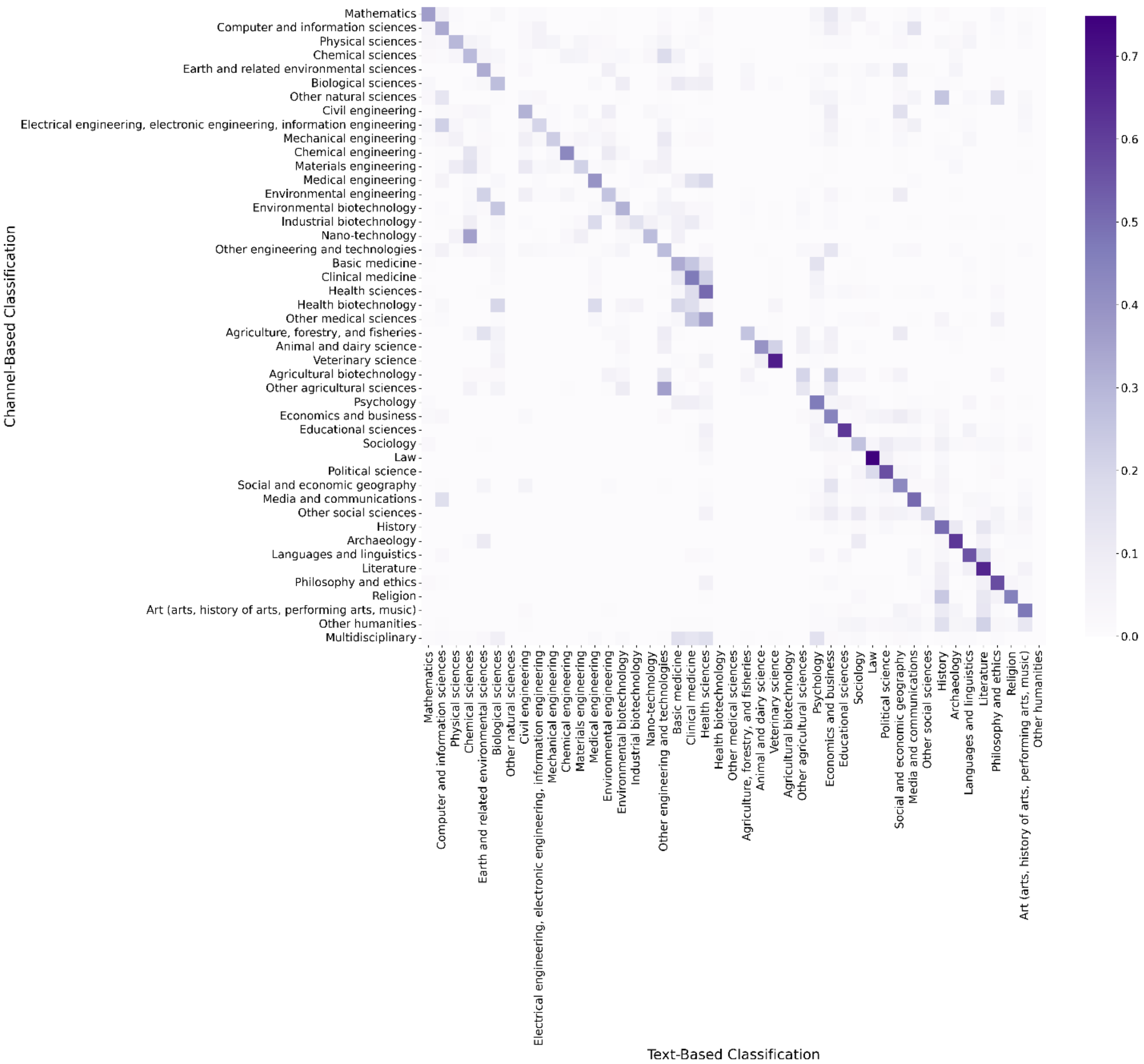

Figure 7. Heatmap of overlap between channel-based and text-based classifications at the discipline level, normalized per row.

At the discipline level (Figure 7), the strongest alignment is typically observed in SSH disciplines, reflecting the origins of the database, though there are exceptions such as “Veterinary science”. Some categories, however, show weaker alignment. “Other Social sciences” is dispersed across a broad set of Social sciences disciplines, which is expected given its nature as a residual category. Similarly, “Sociology” spreads across multiple Social sciences disciplines rather than aligning closely with its own channels.

Several cases point to disciplinary contributions across fields. “Law” shows a significant contribution to “Political science”, consistent with the relevance of legal perspectives for political science research. “History” contributes to multiple Humanities channels, illustrating the broad use of historical framing across humanistic research. “Economics and business” is another notable case, contributing a noticeable share across a wide range of channel-based disciplines, even if not dominant in any of them.

Overall, the results show that SSH-affiliated publications are present in channels across the scientific landscape. The alignment between channel-based and text-based classifications is generally high, but contributions from SSH also extend into many other fields. These patterns reflect the specific profile of the VABB-SHW dataset.

## Comparison of the three classifications

The Jaccard Index is a common metric to assess similarity between sets, chosen here for its simplicity. It is defined as the size of the intersection divided by the size of the union of two sets:

$$J(A,B) = \frac{|A \cap B|}{|A \cup B|}$$

Values range from 0 (no overlap) to 1 (complete overlap). Because publications can be assigned to more than one area or discipline, the Jaccard Index is well suited here: it compares the overlap of the sets of labels assigned by two classifications relative to the total number of labels assigned. This allows us to move beyond descriptive comparisons and perform a quantitative evaluation of the degree of alignment between classification systems at the publication level.

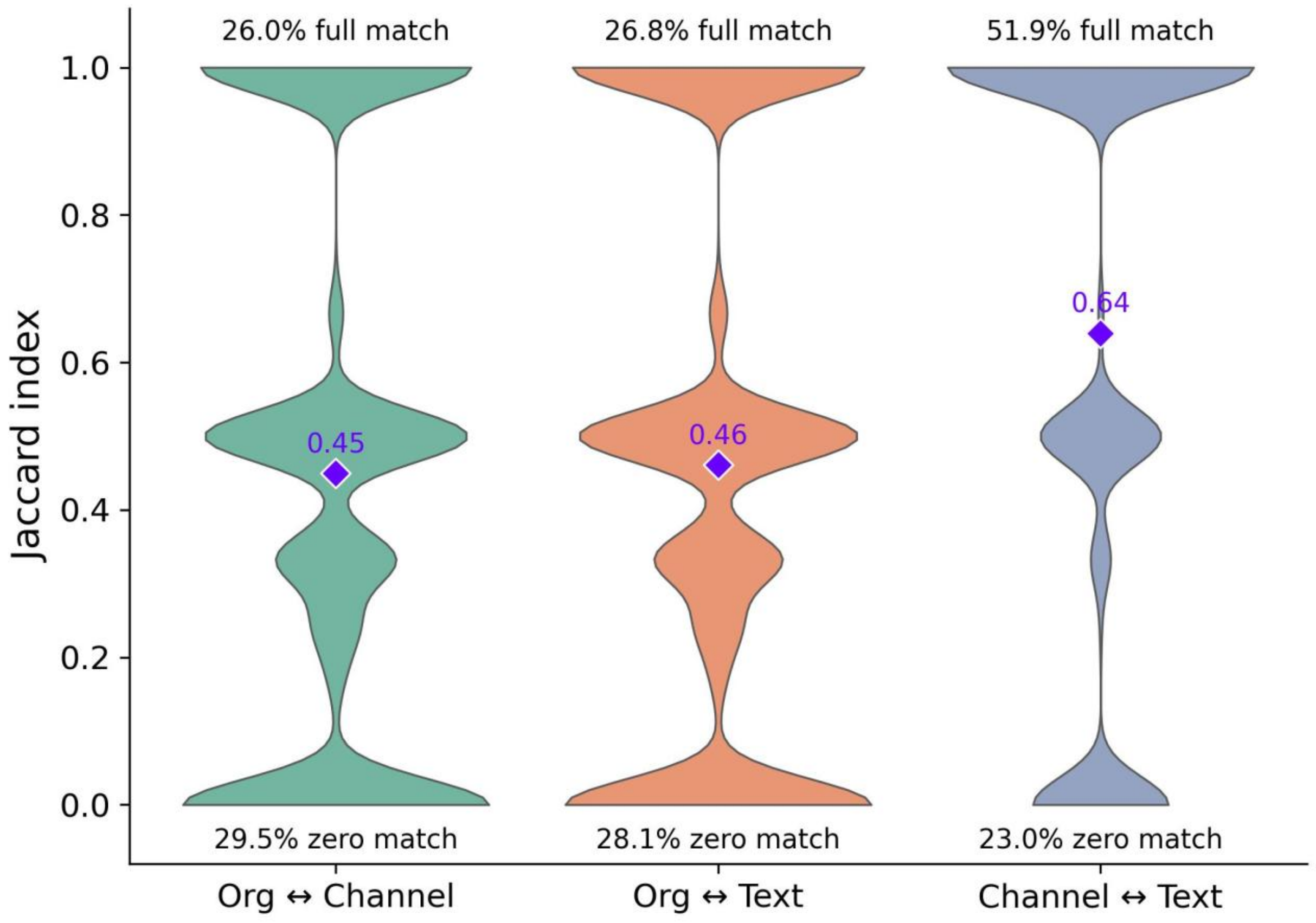

Figure 8. Violin Jaccard Index values for three pairwise comparisons (organizational ↔ text-based, organizational ↔ channel-based, channel-based ↔ text-based). The violins show the full distribution of values across publications and the diamonds mark the mean.

Figure 8 shows the distributions of Jaccard Index values for the three pairwise comparisons. Each plot combines a violin, indicating the shape of the distribution, with a mean marker (diamond). The results confirm that the strongest alignment is between channel-based and text-based classifications. The overlap between organizational and text-based classifications is somewhat higher than that between organizational and channel-based, although the difference is small. The distributions are wide, meaning that alignment varies substantially across individual publications, from cases with no shared categories to cases of complete overlap. Full overlap occurs far more often for channel–text than for the other two comparisons, underscoring the closer relation between these two systems.

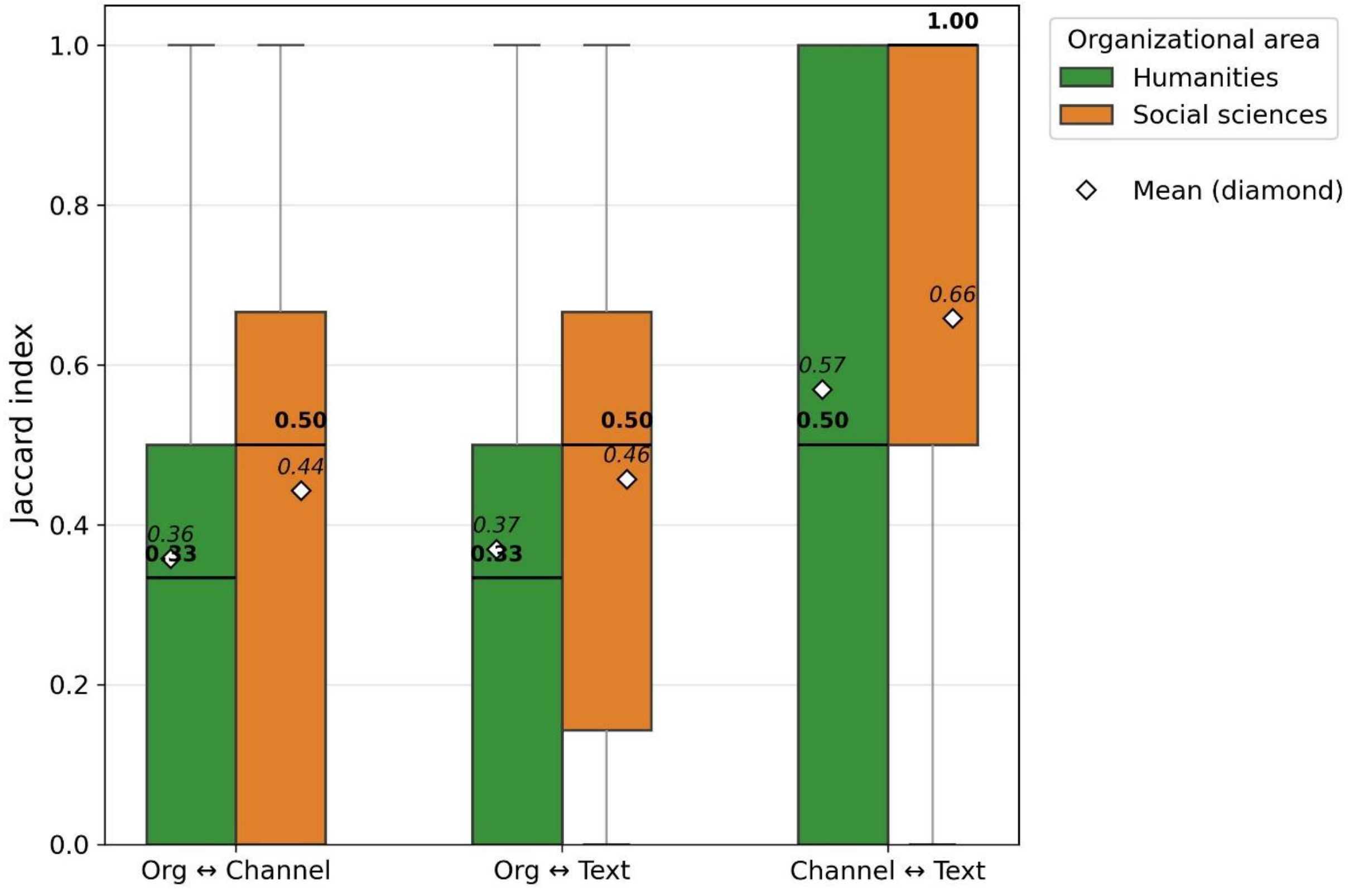

Figure 9. Boxplots of Jaccard Index values for Humanities and Social sciences, comparing organizational, channel-, and text-based classifications. Lines mark the median and diamonds the mean.

Figure 9 presents the same results separately for Humanities and Social sciences. Here again, the highest similarity is found between channel- and text-based classifications. For both areas, organizational-text alignment and organizational-channel are close in both means and medians, with organizational-text alignment overall slightly higher. The spread of values shows that, in both areas, publications range from no overlap to strong agreement, echoing the overall distributions in Figure 8. Taken together, these findings are consistent with the results of the previous subsections: the text-based classification aligns most with the channel-based one, while also showing a closer relation to the organizational classification than the channel-based does.

## Where do organizational and channel-based classifications diverge?

This section explores the set of publications where there is no intersection between the organizational and channel-based classifications, which represents 39,555 publications (26% of the total). The aim is to examine whether the text-based classification provides additional insight into how the content of these publications aligns with the authors' disciplinary affiliation and the orientation of the publication channel.

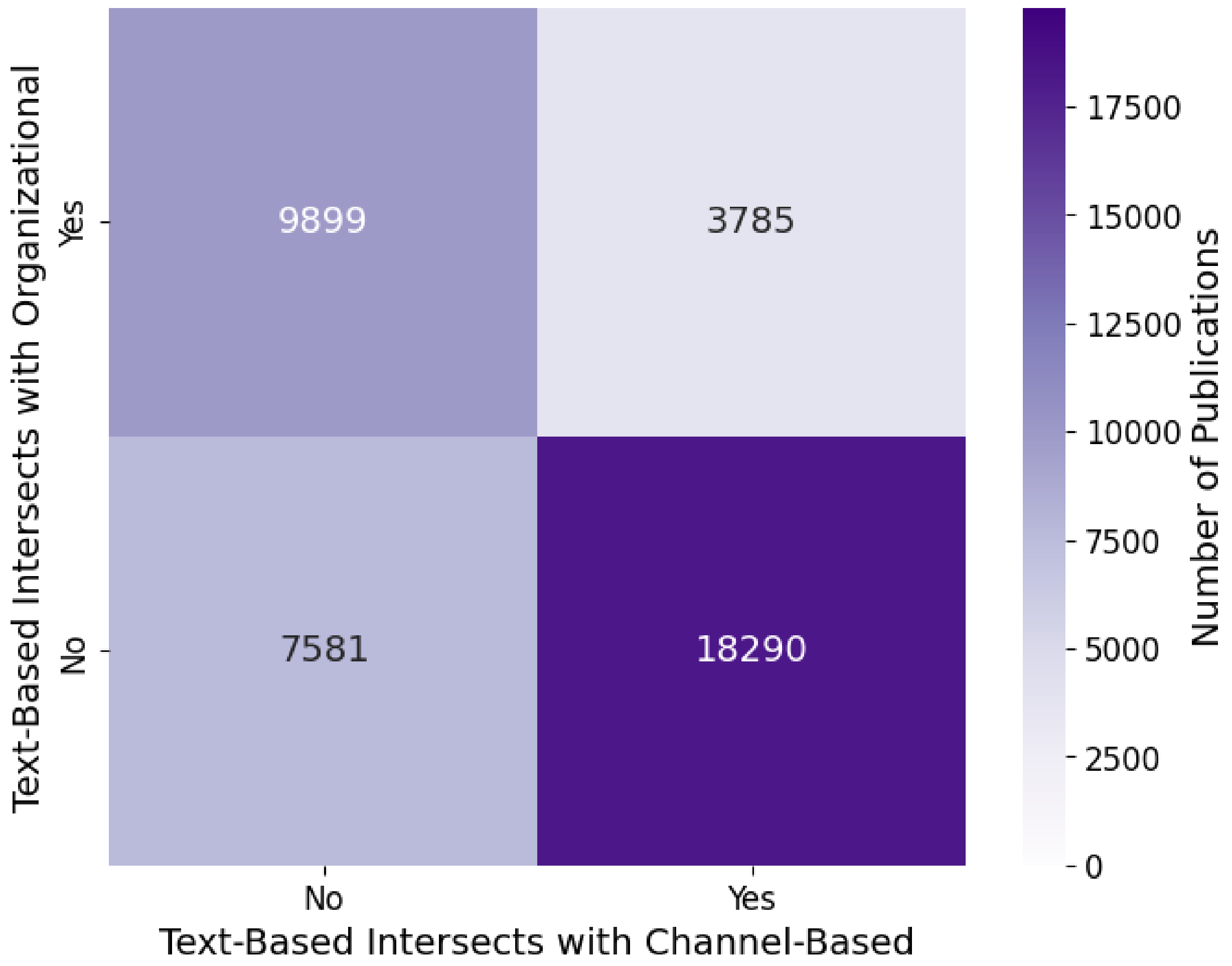

Figure 10. Distribution of alignment of the text-based classification with organizational and channel-based classifications among the 39 555 publications where organizational and channel classifications do not overlap.

Figure 10 shows how the text-based classification distributes across the different alignment options. Most cases align with one of the two systems, and more often with the channel-based classification than with the organizational one. In a smaller share of publications, the text-based classification overlaps with both, effectively bridging the two perspectives. Finally, in about one-fifth of cases, the text-based classification introduces a third disciplinary view, diverging from both the organizational and channel-based classifications.

Sometimes the text-based classification follows the authors' organizational affiliation. For example, in *European Taxation*, the article *"Why the ECJ Should Interpret Directly*

*Applicable European Law as a Right to Intra-Community Most-Favoured-Nation Treatment"* deals with legal aspects of European tax law. It was classified textually as "Law," consistent with the organizational classification, even though the journal itself is categorized under "Economics and business." In other cases, the text-based classification reflects the orientation of the outlet. In *Acta Clinica Belgica*, the article *"Kosteneffectiviteit van vaccinatie tegen pneumokokkenbacteriëmie bij bejaarden: resultaten voor België"* examines the cost-effectiveness of pneumococcal vaccination for elderly people in Belgium. It was classified based on the available textual data as "Health sciences," in line with the medical focus of the journal, while the organizational affiliation placed it under "Economics and business." There are also situations where the text bridges both perspectives. In *Explorations in Financial Ethics*, the article *"Efficiency and rationality in financial markets"* analyzes the functioning of financial markets. The text-based classification assigned both "Economics and business" and "Philosophy and ethics," aligning with the organizational classification in the first case and with the channel-based classification in the second. Finally, there are cases where the text-based classification introduces a completely different perspective. In *Grammaire et enseignement du français (1500–1700)*, the article *"Peeter Heyns, a 'French schoolmaster'"* discusses a sixteenth-century French schoolmaster and his work. The organizational classification assigned it to "Literature," reflecting the author's departmental affiliation, while the channel is categorized under "Languages and linguistics." The text-based classification, however, placed it in "History," based on the available textual information.

The distribution of these publications also varies by publication type (Figure 11). Journal publications dominate the full dataset, but their share decreases in the no-overlap subsets, where book chapters and books represent a greater share. Books are usually more coherent publication channels than journals or conference proceedings, which accommodate a wider variety of contributions. One might therefore expect them to show a higher alignment with text-based classification than journals, which stands in contrast to the results presented in Figure 11.

The picture that emerges through a qualitative analysis is, however, more complex. Books often address a subject from multiple perspectives, and this diversity is reflected in their chapters. For example, many books include a historical overview chapter regardless of the main topic. Other combinations also appear, such as the chapter "*New media, new movements? The role of the internet in shaping the 'antiglobalization' movement"*, which was classified under "Media and communications" while its book, "*Cyberprotest"*, was classified under "Sociology". Further complications arise because the model is trained primarily on journal articles, which may give greater weight to titles than to channels, and because many book chapters lack abstracts, it classifies on limited information. In addition, channel and text classifications sometimes diverge in semantically close areas – for instance, the chapter "*Grammatical theory in Aristotle's Poetics, chapter XX*" from "*Grammatical theory and philosophy of language in Antiquity*" has its channel classified under "Languages and linguistics" while its text-based classification is "Literature". Finally, some cases likely reflect outright misclassifications on either side.

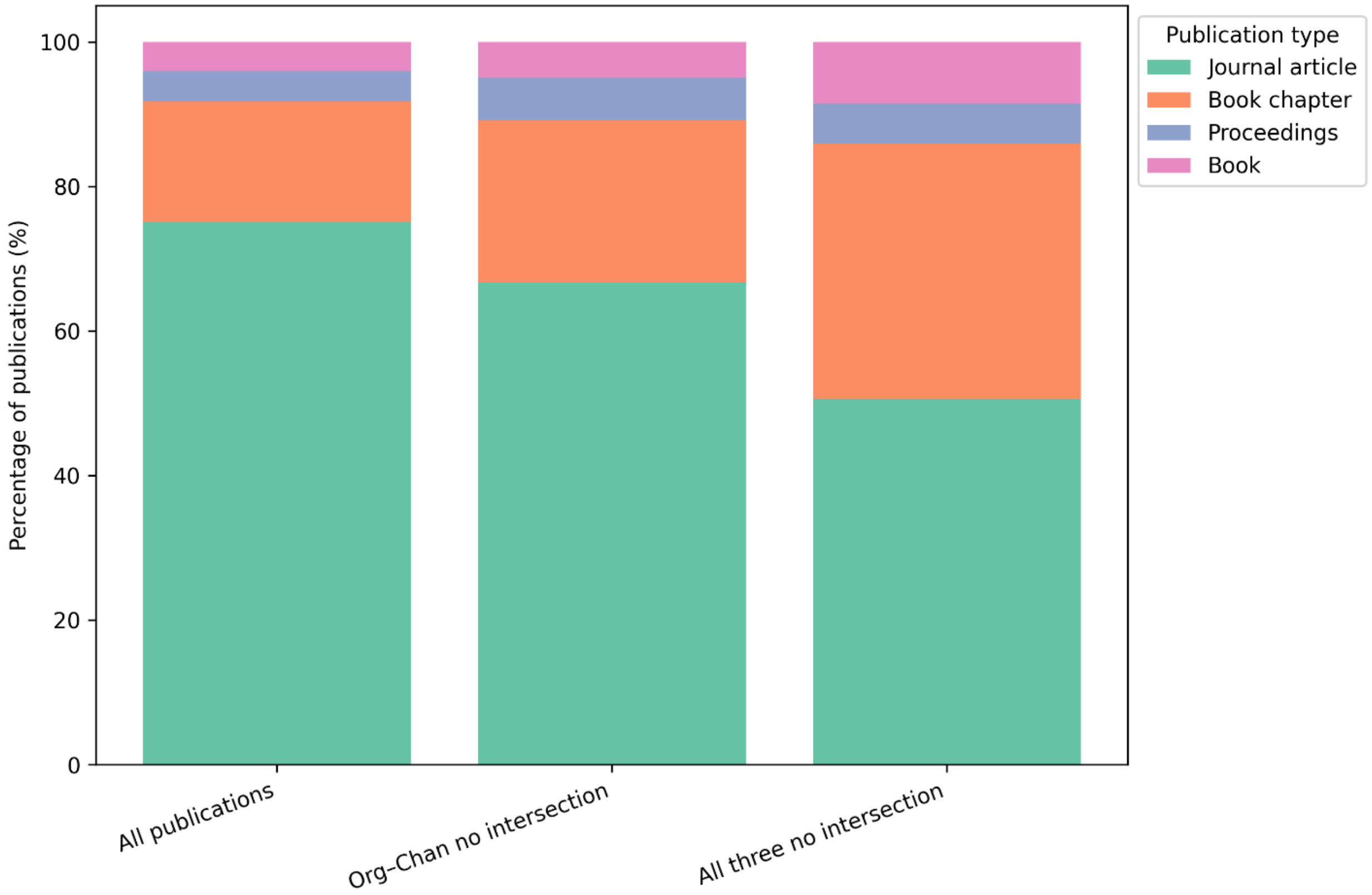

Figure 11. Distribution of publication types across all publications, those without intersection between organizational and channel classifications, and those without intersection across all three classifications

In sum, when organizational and channel-based classifications diverge, the text-based classification tends to align with one of the two, more often with the channel, while also serving at times as a bridge or offering a completely different perspective. These divergences are not limited to particular disciplines but are also linked to publication types, with books and book chapters more frequent among the cases where classifications do not overlap.

## Channel profile comparison: channel labels vs. text-based aggregates

In this section, we shift from individual publications to the channel level. We compare the disciplinary labels assigned to channels with profiles derived from aggregating the text-based classifications of the publications they contain. The analysis covers only outputs from Flemish SSH departments included in VABB-SHW, so it does not capture the full output of these channels. Still, it shows how the contributions of Flemish SSH scholars map onto different channels and the extent to which they align – or diverge – from the channels' assigned disciplinary scope.

## Multidisciplinary channels

Multidisciplinary journals, as mentioned in the introduction, are a common issue of channel-based classifications, since the label obscures the scope of the journal, which limits the conclusions one can draw from channel-based classifications. Here we examine what publication-specific text-based classification can reveal about the publications in these journals. Moreover, looking specifically at Multidisciplinary journals with more than 50 publications in our database, we analyze if they are distributed across a large variety of disciplines or mostly focused on specific ones.

In total, our database contains 73 sources classified as Multidisciplinary: 67 journals, five books and one book series accounting for 1772 publications. Only four Multidisciplinary channels (all journals) have more than 50 publications associated, accounting for a total of 1 339 publications (75.5% of the publications in Multidisciplinary channels). Our analysis therefore concentrates on these four journals.

To contextualize these findings, we also examined the disciplinary distribution of the same journals in WoS. The WoS distributions are based on meso-level citation topics, which were manually mapped to OECD FORD categories using the WoS Science Category to OECD FORD official mapping as reference. In some cases, judgement calls were necessary, typically by consulting higher-granularity topics associated with the meso-level categories.

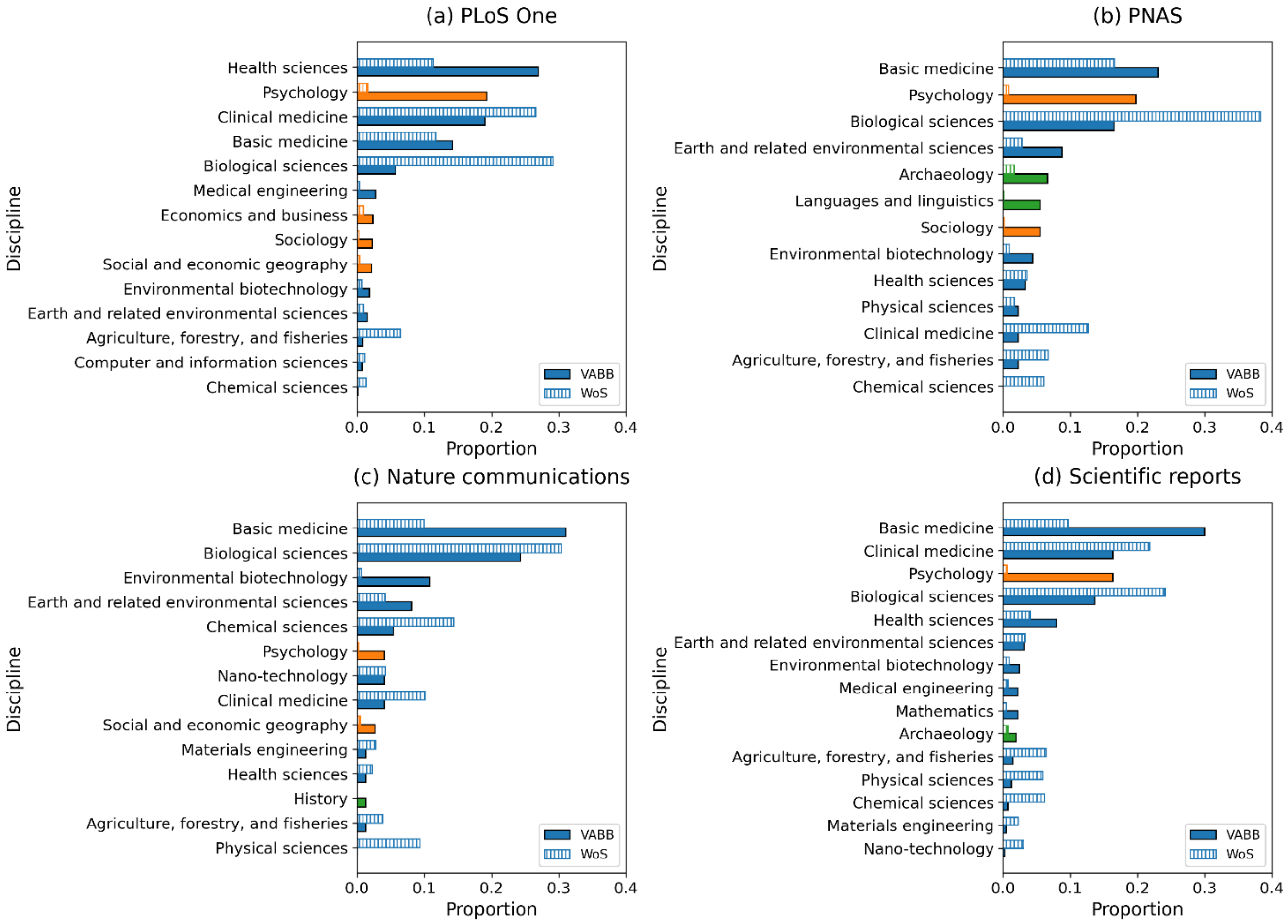

Figure 12. Distribution of disciplines in multidisciplinary journals with more than 50 publications, comparing VABB-SHW text-based classifications and WoS. Bar colors indicate areas: orange = Social sciences, green = Humanities, blue = non-SSH disciplines. The figure includes the combined set of the ten most frequent disciplines from VABB-SHW and the ten most frequent disciplines from WoS for these journals.

Figure 12 shows the distribution of disciplines in the multidisciplinary journals with more than 50 publications, comparing VABB-SHW text-based classifications with WoS. The figure includes the combined set of the ten most frequent disciplines from each source. Medical and Health sciences dominate in both datasets. In VABB-SHW, however, “Basic medicine” and “Health sciences” occur more frequently, while “Clinical medicine” is less represented than in WoS. Social Sciences and Humanities also appear more prominently in VABB-SHW than in WoS, with “Psychology” being one of the main contributing areas. Environmental-related sciences are likewise more visible in VABB-SHW, whereas most Natural sciences and Agricultural sciences are underrepresented compared to WoS.

In addition to journals, VABB-SHW also includes five books and one book series classified as multidisciplinary. The books are represented by only one or two chapters each, and the series (*Springer Handbooks*) by a single title, which provides too little material for further analysis.

## Other channels

Another dimension to explore through this research is the relationship between the classification of the channels and the classification of the publications in those channels. To allow a more comprehensive view, the section focuses on the channels not labelled with the Multidisciplinary label and that have at least fifty publications in our database. There are 336 journals respecting these criteria, having in total 38 015 publications.

To compare channel- and publication-level classifications, we examine whether the disciplinary labels assigned in a channel's existing classification are supported by the aggregated text-based classification of its publications. We define a channel discipline as *supported* if the share of the channel's publications classified in this discipline exceeds the chosen threshold (10%, 20%, … 90%). This does not imply exclusivity: additional disciplines not assigned to the channel may also pass the threshold. The key criterion is simply whether the channel's assigned disciplines are among those that do. Based on this definition, we distinguish three cases: full support (all disciplines from the channel's existing classification are supported), partial support (some but not all are supported, which is only possible when a channel has at least two assigned disciplines), and no support (none are supported).

Figure 13 shows the distribution of journals across these categories as the threshold increases. At the lowest threshold (10%), 289 journals (86%) show full support, while only a small number have no support. At the highest threshold (90%), only 104 journals (31%) still show full support, while the majority (189 journals, 56%) have no support. The sharpest decline occurs between the 80% and 90% thresholds, when 66 journals move from having their classification fully or partially supported to having none of the disciplines supported. At the 90% threshold, about one third of the journals in VABB-SHW behave as truly disciplinary outlets, while most publish a more diverse set of papers when considering the contributions of Flemish SSH scholars.

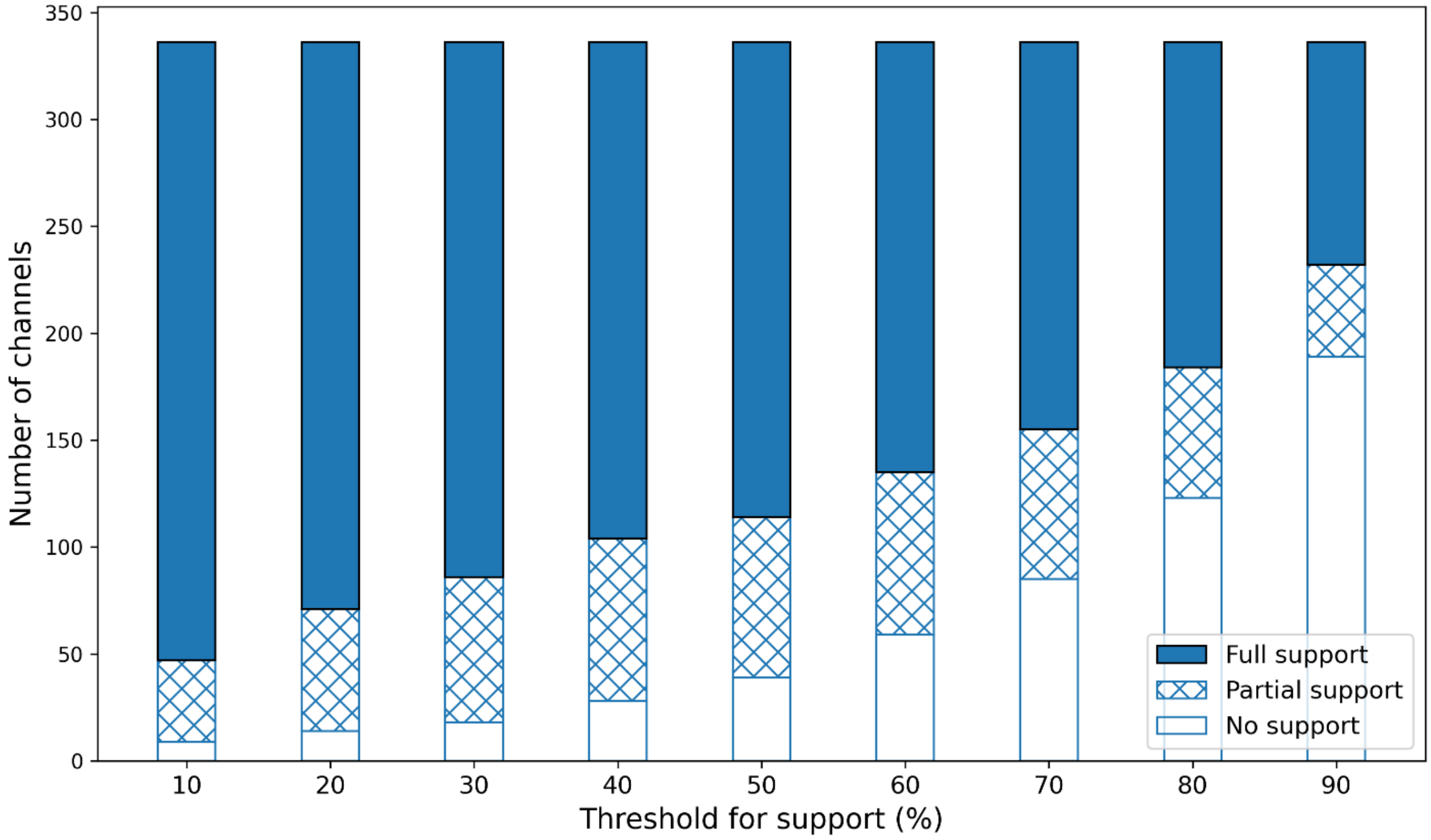

Figure 13. Distribution of journals with full, partial, or no support between their channel-based classification and the publication-level classification, across varying thresholds for considering a journal-level discipline as supported.

To further clarify the role of channel classifications, Figure 14 shows results restricted to single-discipline journals (n = 218). Here only full or no support is possible. At a 10% threshold, 209 journals (95.9%) are supported by their publication-level classifications, but support decreases as thresholds rise: at 90% only 99 journals (45.4%) still show support. These values are significantly higher than for the full set of journals. Since 75.4% of the channels in this analysis are single-labeled, this suggests that most of the publications showing full support belong to single-discipline journals. At the same time, the drop in support at higher thresholds indicates that even journals classified as single-discipline often contain a more diverse set of publications when examined at the publication level.

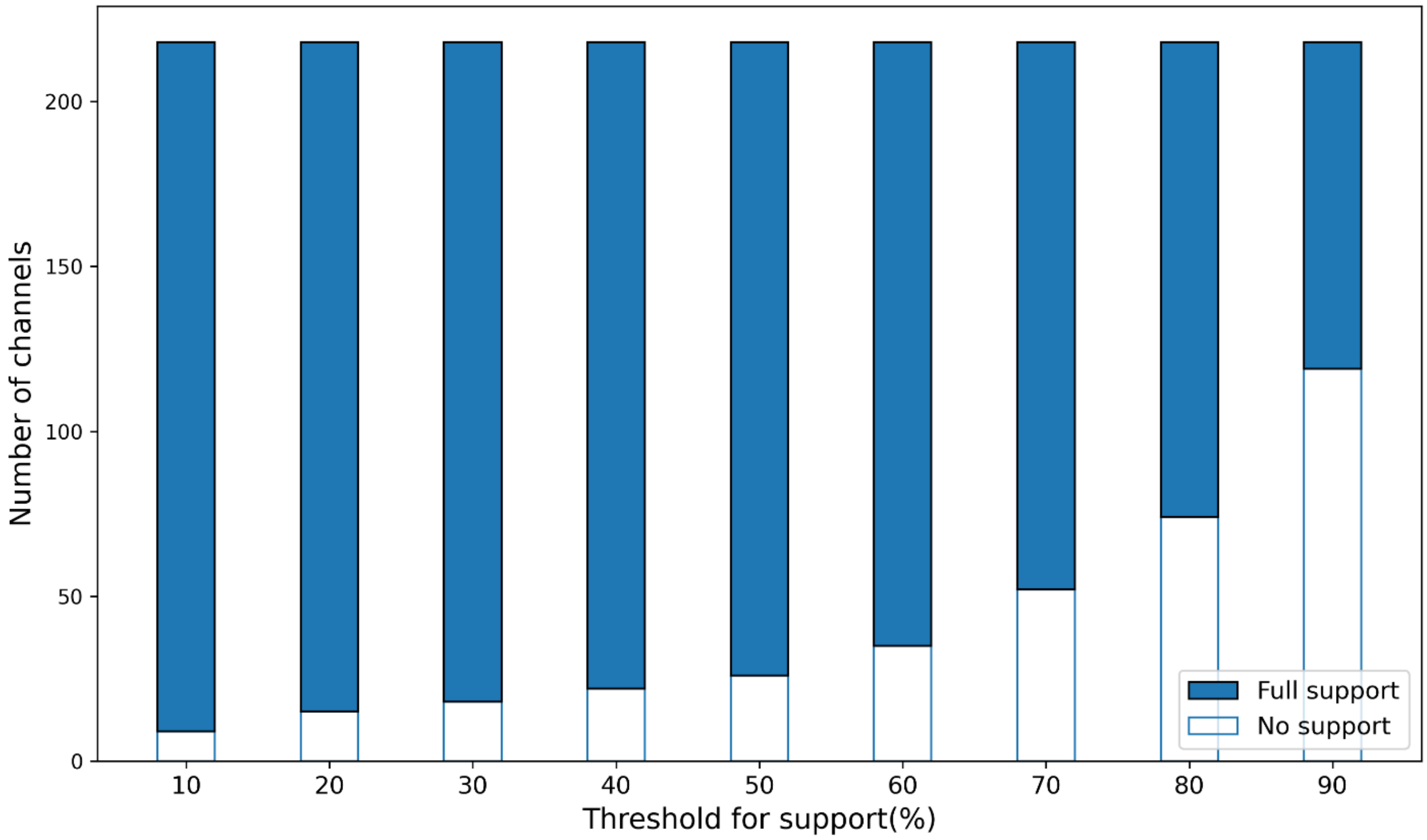

Figure 14. Distribution across thresholds of single-discipline journals with and without support for their disciplines.

A closer look at the journals with partial or no support at low thresholds reveals three distinct situations. First, text-based classification can flag potentially incorrect channel-level classifications, where the assigned discipline does not correspond to the actual content of the journal. For example, *Thema: tijdschrift voor hoger onderwijs en management* (the subtitle translates to "journal for higher education and management") is classified as "Economics and business", while its publications are mainly classified as "Educational sciences". Second, some mismatches reflect systematic effects of the WoS → OECD FORD mapping used in model training: for instance, "Psychology, Psychoanalysis" is mapped to "Health sciences". This propagates into the publication-level assignments and explains mismatches in some psychology journals focused on psychoanalysis, where the channel label points to "Psychology" but the publication-level classification points to "Health sciences". Third, some journals genuinely have a diverse disciplinary profile, such as *Afrika Focus*, which is classified under "Other humanities" and "Other social sciences" but whose publications are spread across "Political science", "Educational sciences", and "Sociology".

Overall, this comparison shows that combining channel-based and publication-based classifications helps validate many existing assignments, but it also highlights cases where channel classifications may be incorrect, cases where systematic effects, such as those of the WoS → OECD FORD mapping, create mismatches, and journals that have a genuinely diverse disciplinary profile. Extending this approach to higher thresholds could provide further insight into how many journals remain supported and may also inform the design of other journal classification systems.

## Discussions and conclusions

This study compared three approaches to disciplinary classification in the VABB-SHW database: organizational, channel-based cognitive, and text-based cognitive. The findings show that the text-based approach often aligns more closely with the channel-based system, indicating that channel classifications do provide relevant insight into the content of publications. At the same time, the text-based classification is closer to the organizational classification than the channel-based classification is, suggesting that the textual features of a publication reflect the authors' disciplinary affiliations more strongly than the choice of channel does. When organizational and channel-based classifications diverge, the text-based classification frequently aligns with one or the other, and sometimes with neither. In a few cases, it intersects with both, effectively acting as a bridge between the authors' organizational discipline and the scope of the publishing channel. It is worth noting, however, that such bridging cases occur only when multiple labels are assigned at the text-based level – a pattern that, according to Figure 2, represents only a small share of publications.

The addition of publication-level classification provides the opportunity to analyze the contributions towards multidisciplinary channels. In our case only 4 multidisciplinary journals passed the threshold of 50 publications deemed relevant for concluding. While channel-based classification treats these journals as single, coherent entities, publication-level assignments reveal the heterogeneity of their contributions. In VABB-SHW, the distribution of disciplinary assignments differs from global patterns for these journals, with a stronger focus on SSH, "Environmental biotechnology", "Basic medicine" and "Health sciences" and a relatively weaker representation of "Clinical medicine" and "Natural sciences". These results align with the SSH focus of the database.

Beyond multidisciplinary journals, publication-level classification also proves useful for assessing contributions to other channels. It can function as a flagging tool by identifying instances where the channel-level classification is not supported by the aggregate of publication-level classifications. The implications are multiple. Misalignments can reveal errors in either classification system, providing opportunities to improve their accuracy. In other cases, they may highlight that the Flemish contributions to a journal mainly cover one particular aspect of its scope. They can also signal shifts in the orientation of a journal over time or the breadth of its disciplinary coverage. At the journal level, fewer than half of outlets with more than 50 publications have their channel-level classification fully or partially supported by more than 90% of publications included in them. This indicates that most channels contain a diverse mix of disciplines, but their classifications remain informative: once the threshold is lowered to 70%, around three quarters of journals with over 50 publications have their channel-level classification fully or partially supported by publication-level assignments.

Certain overall tendencies stand out. Channels with full support are most often those that carry a single disciplinary label, suggesting that narrower-scope channels more consistently contain publication within their designated discipline. Another notable finding is the "interdisciplinary" position of "Sociology". In VABB-SHW, Sociology is often tightly related to other SSH areas, echoing earlier observations in Eykens et al. (2022, 2023) and resonating with Abbott's reflections on the broad coverage of "Sociology" (Abbott, 2001). A similar pattern is visible for "History", which appears in channels across multiple Humanities disciplines and in the outputs of researchers from diverse Humanities fields. Likewise, publications classified as "Economics and business" are found in channels across a wide range of disciplines, including those outside SSH. By contrast, publications classified as "Law" are concentrated in "Law" channels (with some presence in "Political sciences"), authored predominantly by researchers from "Law" and "Criminology".

Taken together, these findings show the value of comparing classification perspectives. Text-based, publication-level classification is not a final authority, but it makes visible both alignments and misalignments that would otherwise remain hidden. For research evaluation and database management, this perspective offers practical benefits: it can validate existing classifications at an aggregate level, flag systematic inconsistencies, and reveal how multidisciplinary journals contribute across fields. For the study of disciplines, it provides empirical evidence of how boundaries overlap and shift, underlining that classifications are best understood as approximate representations rather than fixed truths.

## Limitations and Future Work

This study has several limitations. At the conceptual level, disciplinary boundaries are inherently fuzzy, which makes precise classification impossible. Disciplines overlap, evolve, and are open to multiple interpretations, meaning that different but equally valid perspectives can exist for classifying the same publication, journal or research group. Results are also shaped by the reference system chosen: the OECD-FORD categories as adapted in VABB-SHW provide a particular lens, and using another framework might produce somewhat different insights.

The analysis also reflects characteristics of the underlying data. Because it is based on the Flemish VABB-SHW database, its conclusions cannot be assumed to generalize directly to other national or international databases. Coverage matters: for instance, only four multidisciplinary journals passed the 50-publication threshold applied here. They provide insightful case studies, but the generalizability of those insights to multidisciplinary journals in other SSH databases is limited. More generally, while multiple channel types are included in VABB-SHW, the requirement of 50 publications per channel restricted the comparisons to large journals with sufficient articles in VABB-SHW, leaving smaller outlets unstudied.

The three classification systems compared here are not perfect. Each contains errors, and although this approach helps flag and correct some of them, others inevitably affect the results. Our observations suggest that the extent of such errors is not sufficient to alter the overall patterns at the aggregate level, but they remain part of the picture. In addition, methodological choices shape the classifications. For channel-based assignments, these choices derive from the external databases used and the way their categories are combined. For the text-based classification, they involve the definition of ground truth, the selection of training data, and the probability threshold used to assign a discipline to a publication. Future work could investigate the sensitivity of results to these parameters.

Finally, the analysis treats both channels and disciplines as static, whereas in reality they evolve over time. Shifts in journal scope or disciplinary vocabularies may change the alignment between organizational, channel-based, and text-based classifications, and these dynamics are not captured here.

Beyond methodological refinements such as testing parameter choices or incorporating measures of classification certainty, future work could apply the same comparative framework to other national databases. This would help clarify which findings are specific to VABB-SHW and which reflect broader dynamics in the SSH environment.

## Contributions

Conceptualization – Cristina Arhiliuc, Tim Engels; Methodology – Cristina Arhiliuc, Tim Engels; Data curation – Cristina Arhiliuc; Formal analysis and visualization – Cristina Arhiliuc, Tim Engels; Writing original draft – Cristina Arhiliuc; Writing review and editing – Cristina Arhiliuc, Raf Guns, Tim Engels; Supervision – Raf Guns, Tim Engels; Funding acquisition – Tim Engels.

## Declaration regarding the use of AI

AI technology was used as an assistant in this study under the direct supervision of the authors. Specifically:

- **Writing**: ChatGPT-5 was used to help structure ideas, refine phrasing, and edit drafts. A question-answer exchange format (ChatGPT-5 → authors) was employed to brainstorm, ensure completeness, and improve clarity.
- **Coding**: Gemini (via Google Colab) and ChatGPT-5 were used as coding assistants, primarily to generate scripts for data processing and visualization. All code was checked, adapted, and validated by the authors.

Final interpretations, decisions, and conclusions were made solely by the authors.

Zhang, J., & Shen, Z. (2024). Analyzing journal category assignment using a paper-level classification system: Multidisciplinary sciences journals. *Scientometrics*. https://doi.org/10.1007/s11192-023-04913-0